\documentclass[a4paper]{jpconf}
\usepackage{graphicx}
\usepackage{float}
\usepackage{psfrag}
\usepackage{cite}

\begin{document}

\title{Polarization of Cosmic Microwave Background}

\author{A Buzzelli$^{1,2}$, P Cabella$^{1,2}$, G de Gasperis$^{1,2}$ and N Vittorio$^{1,2}$}

\address{$^{1}$ Dipartimento di Fisica, Universit\`a di Roma ``Tor Vergata'', via della Ricerca Scientifica 1, I-00133, Roma, Italy}
\address{$^{2}$ Sezione INFN Roma 2, via della Ricerca Scientifica 1, I-00133, Roma, Italy}

\ead{alessandro.buzzelli@roma2.infn.it}

\begin{abstract}
In this work we present an extension of the ROMA map-making code for data analysis of Cosmic Microwave Background polarization, with particular attention given to the inflationary polarization B-modes. The new algorithm takes into account a possible cross-correlated noise component among the different detectors of a CMB experiment. We tested the code on the observational data of the BOOMERanG (2003) experiment and we show that we are provided with a better estimate of the power spectra, in particular the error bars of the BB spectrum are smaller up to 20\% for low multipoles. We point out the general validity of the new method. A possible future application is the LSPE balloon experiment, devoted to the observation of polarization at large angular scales. 
\end{abstract}

\section{Introduction}

The Cosmic Microwave Background (CMB) is the relic radiation of the early universe. It is characterized by an almost perfect black body spectrum, with average temperature of about 2.7 K. It is extremely isotropic with temperature fluctuations of only few parts out of $10^{5}$. Nonetheless, the temperature anisotropy pattern has been a fundamental tool to extract cosmological information, since it succeeded in constraining cosmological parameters with very high accuracy. However, to break the degeneracy among some parameters we need to study the (linear) polarization pattern of the CMB, due to Thomson scattering of quadrupolar anisotropies at last scattering surface. Since the polarization is characterized by an amplitude at least one order of magnitude lower than temperature anisotropy, its observation has been an arduous experimental challenge. \\Polarization can be decomposed into a gradient component (E-modes) and a curl component (B-modes). While E-modes have been widely observed and analyzed, we still do not have a unique evidence of primordial B-modes. The search for polarization B-modes is perhaps the most exciting and promising field in modern cosmology. We know that scalar (density) perturbations produce only E-modes, while tensor perturbations, i.e. Gravitational Waves (GW), produce both E- and B-modes. Therefore, a detection of B-modes would provide a definitive confirmation to the existence of a stochastic GW background, as predicted by Inflation theory.
\\Numerous experiments have been done to observe the microwave sky, from ground, balloons and space, each one characterized by its own scanning strategy and experimental setup. In this work we focus on two balloon missions: BOOMERanG (2003) and LSPE, which will be likely launched in 2016.

\section{CMB polarization: theory}

In this section we briefly review the theoretical and mathematical aspects of CMB polarization. We follow mainly the formalism of \cite{CK}. A complete treatment of polarization can be achieved by the two Stokes parameters $Q$ and $U$, since $I$ describes temperature only and $V$ vanishes, because Thomson scattering induces no circular polarization. We define the polarization tensor $P_{ab}$ as:

\[ P_{ab}(\hat{n}) =\frac{1}{2} \left( \begin{array}{cc}
Q(\hat{n}) & -U(\hat{n})sin\theta \\
-U(\hat{n})sin\theta & -Q(\hat{n})sin^{2}\theta \end{array} \right).\]

\noindent where $\hat{n}=(\theta,\phi)$ is the direction in the sky. Note that this tensor is symmetric, $P_{ab}=P_{ba}$, and trace-free, $g^{ab}P_{ab}$, with the metric given by $g_{ab}=diag(1,sin^{2}\theta$).\\
A generalization of the Helmotz theorem states that any symmetric trace-free rank-2 tensor can be decomposed into a gradient and a curl component. In our case, we refer to these components as E-modes and B-modes, respectively. More explicitly, the polarization tensor can be expanded in terms of basis functions that are gradients and curls of tensor spherical harmonics:

\begin{equation}
\frac{P_{ab}(\hat{n})}{T_{0}}=\sum_{l=2}^{\infty}\sum_{m=-l}^{l}[a_{(lm)}^{E}Y_{(lm)ab}^{E}(\hat{n})+a_{(lm)}^{B}Y_{(lm)ab}^{B}(\hat{n})],
\end{equation}

\noindent where $T_{0}$ is the average temperature. The expansion coefficients and the spherical harmonics are given by:

\begin{equation}
a_{(lm)}^{E}=\frac{1}{T_{0}}\int d\hat{n}P_{ab}(\hat{n})Y_{(lm)}^{E\, ab*}(\hat{n}),
\end{equation}

\noindent 
\begin{equation}
a_{(lm)}^{B}=\frac{1}{T_{0}}\int d\hat{n}P_{ab}(\hat{n})Y_{(lm)}^{B\, ab*}(\hat{n}),
\end{equation}

\noindent 
\begin{equation}
Y_{(lm)ab}^{E}=N_{l}(Y_{(lm):ab}-\frac{1}{2}g_{ab}Y_{(lm):c}^{c}),
\end{equation}

\noindent 
\begin{equation}
Y_{(lm)ab}^{B}=\frac{N_{l}}{2}(Y_{(lm):ac}\epsilon_{b}^{c}+Y_{(lm):bc}\epsilon_{a}^{c}),
\end{equation}

\noindent where `:' denotes the covariant derivative, $\epsilon_{ab}=antidiag(\sqrt{g},\sqrt{-g})$ with $g$ being the determinant of $g_{ab}$, and the normalization coefficient $N_{l}$ is:

\begin{equation}
N_{l}=\sqrt{\frac{2(l-2)!}{(l+2)!}}.
\end{equation}

\noindent From the expansion coefficients we introduce the power spectra as:

\begin{equation}
<a_{(lm)}^{X*}a_{(l^{'}m^{'})}^{X^{'}}>=C_{l}^{XX^{'}}\delta_{ll^{'}}\delta_{mm^{'}},
\end{equation}

\noindent where $X,X^{'}=\{E,B\}$. According to the standard cosmological model, $C_{l}^{TB}=C_{l}^{EB}=0$. Recent estimates of $C_{l}^{TT}$, $C_{l}^{TE}$ and $C_{l}^{EE}$ can be found in \cite{Pla}. The new frontier of CMB observation is the search for the (primordial) $C_{l}^{BB}$.\\
We already said that B-modes are univocally produced by tensor perturbations. However, we need to consider two important effects that may introduce a B-component in the polarization: gravitational lensing and foreground contamination. The gravitational lensing is caused by large scale mass inhomogeneities between us and the last scattering surface. Its effect is dominant at large multipoles. Foreground contamination is mainly due to synchrotron radiation, below 60 GHz, and dust emission, above 90 GHz. Therefore, the detection of B-modes needs a careful characterization of these effects, in order to isolate the B-component of inflationary origin.\\
The exact level of B-mode signal is still uncertain and it depends on the Inflation potential. Qualitatively, we have $C_{l}^{TT}\simeq{10C_{l}^{TE}}\simeq{10^{2}C_{l}^{EE}}\simeq{10^{4}C_{l}^{BB}}$.

\section{CMB polarization: data analysis}
The extraction of cosmological information from observations is a complex procedure and it depends on the type of the experiment (ground based, balloons, satellites). However, the general pipeline is the same for any experimental setup and it can be separated into three main steps.\\ An experiment observes the sky and collects data in a given temporal sequence: we refer to these data as TOD (Time Ordered Data). The TOD must be calibrated, i.e. expressed in $\mu K$, and the pointing must be reconstructed. The first step of the pipeline is the projection of the TOD on the sky, that means to build a sky map (map-making). This implies a compression from $10^{10}-10^{9}$ TOD to $10^{7}-10^{6}$ pixels of a map. The map-making is the most time consuming part of the data analysis. Therefore, an important task is to find a computationally efficient way to perform the analysis that preserves all the relevant cosmological information. A detailed description of map-making can be found in \cite{Nat1} and \cite{Nat2}. The second step of the pipeline is the estimation of the angular power spectra from the maps, that implies a compression to $10^{4}-10^{3}$ data. In general, the sky maps contain ``bad'' pixels, i.e. pixel not observed or contaminated. For example, balloon experiments are designed to observe only a fraction of the sky, and even all-sky missions, like Planck \cite{Pla2}, have to remove part of the sky from the analysis, due to foreground contamination. Therefore, we have first to estimate the spectra referred to the maps (pseudo-spectra) and then statistically extend the spectra to the celestial sphere (full sky-spectra) through appropriate algorithms. To do that, in our analysis we used the HEALPix software \cite{Gor} and the MASTER algorithm \cite{Hiv}, respectively. The third step is the computation of the $\simeq10$ cosmological parameters, such as the expansion rate and the curvature of the universe, and the relative abundances of baryons, Dark Matter and Dark Energy, from the spectra. This can be done with the COSMOMC software \cite{Lew}.\\Our study is focused mainly on the first two steps. We know that the detectors of a CMB experiment may be affected by strong cross-correlations, i.e. the noise of each detector is the sum of a non-correlated component and a common-mode component seen by all the detectors \cite{Pat}. Our aim is to test the improvement that we gain in the estimate of the spectra when we take into account the cross-correlated noise component, usually neglected in past works, in the map-making code.\\The TOD from one detector can be placed in a vector $d$, defined as: 

\begin{equation}
d=Pm+n.\label{1}
\end{equation}

\noindent where the operator $P$ is the ``pointing matrix'', containing all the information about the scanning strategy, the vector $m$ is the map and the vector $n$ is the instrumental noise. In general, $m$ refers to the Stokes parameters $I$, $Q$ and $U$. Some assumptions are made for the noise \cite{Pat}, in particular we assume to know all its statistical properties and we take it as stationary. In case of more than one detector, Eq.~(\ref{1}) is still valid, with the quantities from each detector being concatenated end to end.\\Map-making means to solve Eq.~(\ref{1}) for $m$. A good estimator $\tilde{m}$ of $m$ can be found by a Generalized Least Squares approach:

\begin{equation}
\tilde{m}=(P^{T}N^{-1}P)^{-1}P^{T}N^{-1}d.\label{2}
\end{equation}

\noindent where $N$ is the ``noise covariance matrix'' defined as $N=<nn^{T}>$. Eq.~(\ref{2}) can be iteratively solved by the method of conjugate gradients.\\To include cross-correlation, we write the noise of the $i$-th detector at time $t$ as \cite{Pat}:

\begin{equation}
n_{t}^{i}=\tilde{n}_{t}^{i}+\gamma^{i}c_{t},
\end{equation}

\noindent where $\tilde{n}_{t}^{i}$ is the uncorrelated term, $c_{t}$ is the correlation, dependent only on time, and $\gamma^{i}$ is an
amplitude, dependent only on the detector. Generally, for each detector, the spectrum of the noise has the following shape:

\begin{equation}
P(f)=A\left[1+\left(\frac{f_{k}}{f}\right)^{\alpha}\right],
\end{equation}

\noindent where $f$ is the frequency, and $A$, $f_{k}$ (knee frequency) and $\alpha$ are constants. We have then a $1/f$ part at low $f$ and a constant part at high $f$.

\section{BOOMERanG (2003)}
\begin{figure}[H]
\begin{center}
\includegraphics[scale=0.3]{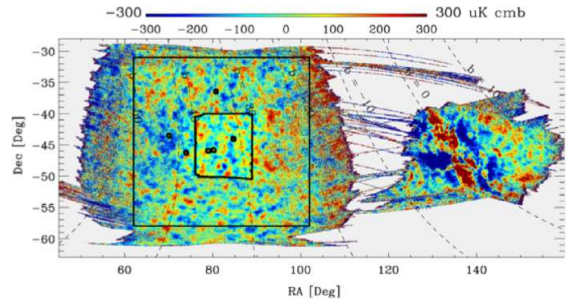}
\end{center}
\caption{Regions scanned by BOOMERanG (2003): the deep region inside the small square, the shallow region inside the big square, and the portion of Galactic plane on the right \cite{Pia}.}
\label{fig1}
\end{figure}

BOOMERanG is a telescope mounted on a stratospheric long duration balloon \cite{Mas, Mac, Mon}, whose main goal is the accurate measurement of the sky in the microwave band. Two missions were launched: in 1998 and in 2003, both from Antarctica. We focus on the second mission, aimed at the polarization detection. The observation strategy was designed to cover a small fraction of the sky with ``deep'' integration to observe polarization, a wider fraction of the sky with ``shallow'' integration to observe temperature, and a portion of the Galactic plane (Fig.~\ref{fig1}).

\begin{figure}[H]
\begin{center}
\includegraphics[scale=0.2]{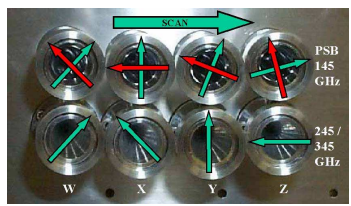}
\end{center}
\caption{Focal plane of BOOMERanG (2003): four pairs of detectors (Polarization Sensitive Bolometers) at 145 GHz and four single detectors at 245 and 345 GHz \cite{foc}.}
\label{fig2}
\end{figure}

\noindent All the survey was carried out in three frequency bands centered at 145, 245 and 345 GHz. For the 145 GHz band, there are four pairs of detectors; both the other bands have four single detectors (Fig.~\ref{fig2}). We concentrate our analysis on the 145 GHz band.

\begin{figure}[H]
\begin{center}
\psfrag{Multipolel}[c][][3]{Multipole $\ell$}
\psfrag{ClBBold}[c][][3]{$\ell(\ell+1) C_{\ell}^{BB}/(2\pi) \hskip 0.8em [\mu \mathrm{K}^{2}]$}
\includegraphics[angle=90,height=4cm,width=6cm]{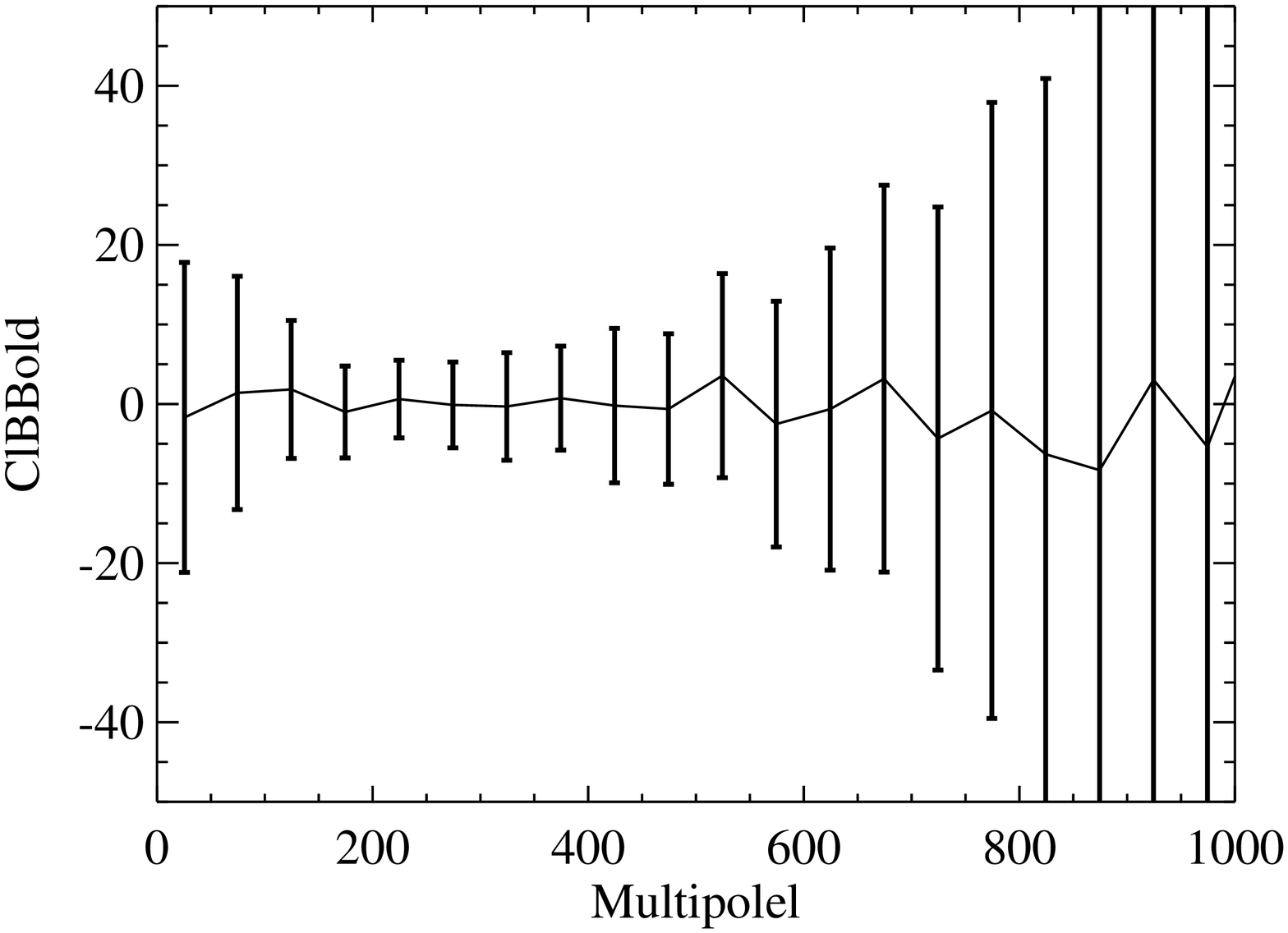}
\psfrag{Multipolel}[c][][3]{Multipole $\ell$}
\psfrag{ClBB}[c][][3]{$\ell(\ell+1) C_{\ell}^{BB}/(2\pi) \hskip 0.8em [\mu \mathrm{K}^{2}]$}
\includegraphics[angle=90,height=4cm,width=6cm]{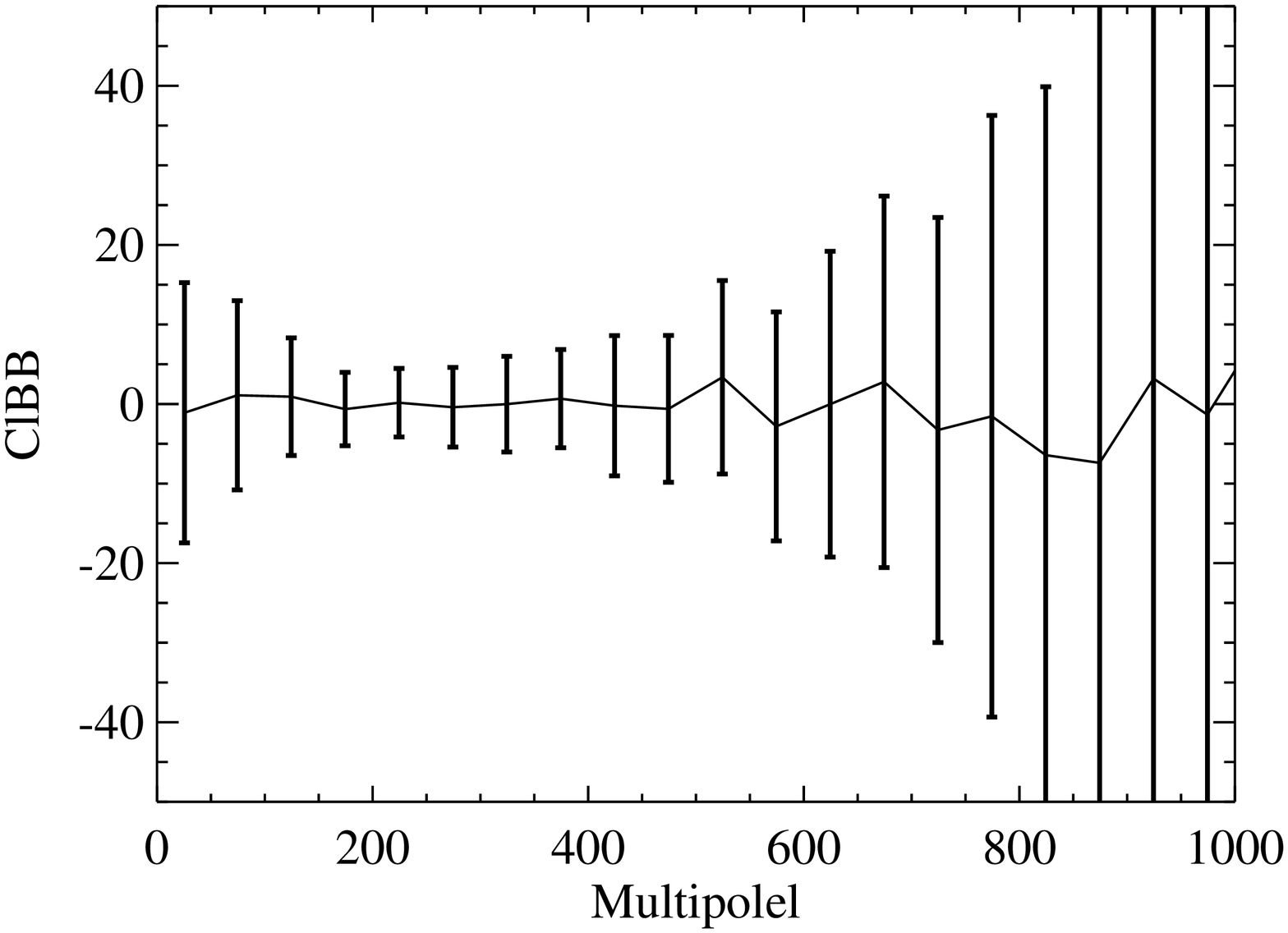}
\end{center}
\caption{Comparison between the BB power spectra estimated neglecting (on the left) and including (on the right) the cross-correlated noise component.}
\label{fig3}
\end{figure}

\noindent We applied the procedure described in the previous section to 200 Monte Carlo simulations of BOOMERanG (2003) data. In particular we used the map-making code ROMA \cite{deG}, extended with a possible cross-correlation among the detectors. To include the effect of the cross-correlated noise component, we added two contributions to the detector auto-noise: the cross-correlated noise between detectors in the same pair and the one among each of the other six detectors. These contributions share with the auto-noise the shape $(1/f)^{2}$ at low $f$, but they are characterized by a constant level at high $f$ of 10\% and 1\% of the auto-noise, respectively.
\\Once generated the maps, we estimated the power spectra. In Fig.~\ref{fig3} we compare the BB power spectrum obtained neglecting and including cross-correlation, using data from the deep region.

\noindent To estimate the effect of considering the cross-correlated noise component, we analysed the ratio of the spectra error bars estimated with and without cross-correlation. We expect that neglecting the cross-correlation will bring to overestimate the noise contribution for low multipoles.

\begin{figure}[H]
\begin{center}
\psfrag{Multipolel}[c][][3]{Multipole $\ell$}
\psfrag{ratio}[c][][3]{Ratio error bars}
\includegraphics[angle=90,height=4cm,width=6cm]{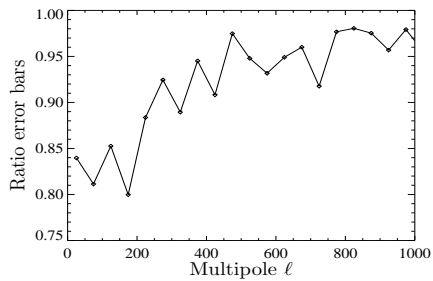}
\end{center}
\caption{Ratio between the BB spectra error bars, when including or not the noise cross-correlation.}
\label{fig4}
\end{figure}

\noindent From Fig.~\ref{fig4} we notice that the new algorithm has provided smaller error bars on the BB spectrum. In particular, the errors are reduced by 20\% for low multipoles. Since the BB spectrum is still buried into instrumental noise, the reduction of the error bars represents a very important step towards a first detection of polarization B-modes.

\section{LSPE}
The techniques shown in last section have general validity. We applied the map-making code to simulated data of BOOMERanG (2003) and estimated the power spectra, showing the improvement obtained by taking into account the cross-correlation among the detectors. The same pipeline can be applied to other microwave experiments, in particular a suitable future application is the LSPE (Large Scale Polarization Explorer) balloon mission \cite{Aio}. Its launch is expected for the end of 2016. It is aimed at observing CMB polarization at large angular scales with a sensitivity about ten times greater than Planck's. The primary target is to constrain the B component of polarization whose information is encoded in the low multipoles. We said that B-modes represent a unique signature of tensor perturbations produced during Inflation. For this purpose, LSPE is expected to improve the limit on the ratio of tensor to scalar perturbations amplitudes down to $r=0.03$. A detection of $r$ would constrain the energy scale of Inflation according to the relation $E=3.3\times10^{16}r^{1/4}GeV$. Moreover, LSPE will produce wide maps of foreground polarization generated in our Galaxy by synchrotron and interstellar dust emission. During its circumpolar Arctic flight, LSPE will observe a large fraction of the northern sky (25\% of the whole sky) with angular resolution of 1.5 degrees FWHM. The payload will host two instruments: an array of bolometric polarimeters, SWIPE \cite{deB}, which will map the sky at 95, 145 and 245 GHz, and an array of coherent polarimeters, STRIP \cite{Ber}, which will survey the same sky region at 43 and 90 GHz. The multi-frequency information provided by LSPE would allow to characterize the foreground contamination, in order to disentangle the single polarized components. Our next step will be the generation of simulated maps in order to estimate the possible improvement that the new algorithm will provide for LSPE. Our interest will be focused on the SWIPE instrument, characterized by two focal planes with a total number of 110 detectors for frequency. We expect that the problem of cross-correlated noise will be a crucial issue in the data analysis of LSPE.

\section{Conclusions} 
In this work we tested a new code for data analysis of CMB polarization, which takes into account a possible cross-correlated noise component among the detectors. We applied this algorithm to simulated data of BOOMERanG (2003) and we showed the benefit obtained in the estimate of the power spectra. In particular, we reduced the BB spectrum error bars up to 20\% at low multipoles. The whole pipeline tested on BOOMERanG can be entirely applied to the data analysis of the forthcoming LSPE mission. Its goal will be mainly the characterization of polarization at large angular scales, where the information of B-modes is encoded. The reduction of the error bars achieved with the new code could be crucial in the search for polarization B-modes with LSPE.

\ack{We thank P. de Bernardis, S. Masi and the BOOMERanG collaboration for providing us with the observational dataset of the BOOMERanG (2003) experiment and K. M. Gorski et al. for the use of the HEALPix package.}

\end{document}